\documentclass{elsart}
\usepackage{epsfig}
\newcommand{\lp}{\left(}
\newcommand{\rp}{\right)}
\newcommand{\lb}{\left[}
\newcommand{\rb}{\right]}
\newcommand{\sigsq}[1]{\sigma^2_{N_{t,k}^{\left( #1 \right)}}}
\newcommand{\sig}[1]{\sigma_{N_{t,k}^{\left( #1 \right)}}}
\begin{document}
\begin{frontmatter}
\title{On the application of differences in intrinsic
fluctuations of Cherenkov light images for
separation of showers.}
\author[Barnaul]{V. V. Bugayov\thanksref{corr_author}}
\author[Barnaul]{A. V. Plyasheshnikov}
\author[FLWO]{V. V. Vassiliev}
\author[FLWO]{T. C. Weekes}

\address[Barnaul]{Department of Physics, Altai State
University, Dimitrova 66, Barnaul, Russia, 656099}
\address[FLWO]{Fred Lawrence Whipple Observatory, Harvard-Smithsonian
CfA, P.O. Box 97, Amado, AZ 85645-0097}

\begin{abstract}
The sensitivity of ground-based imaging atmospheric Cherenkov
$\gamma$-ray observatories depends critically on the primary
particle identification methods which are used to retain
photon-initiated events and suppress the spurious background
produced by cosmic rays. We suggest a new discrimination technique
which utilizes differences in the fluctuations of the light
intensity in the images of showers initiated by photons and those
initiated by protons or heavier nuclei. The database of simulated
events for the proposed VERITAS observatory has been used to
evaluate the efficiency of the new technique. Analysis has been
performed for both a single VERITAS imaging telescope, and a
system of these telescopes. We demonstrate that a discrimination
efficiency of $\ge 1.5 - 2.0$ can be achieved in addition to
traditional background rejection methods based on image shape
parameters.

\end{abstract}


\thanks[corr_author]{Corresponding author:
bugaev@theory.dcn-asu.ru}

\end{frontmatter}

\section{Introduction}

Gamma-ray astronomy in the energy range above 100 GeV has made
dramatic progress in the last decade due to rapid development of the
atmospheric Cherenkov technique at a number of ground-based $\gamma$-ray
observatories.  Detection of Cherenkov radiation from atmospheric
cascades initiated by high energy particles has been considered as a
potentially promising method for high energy $\gamma$-ray astronomy
since the fifties~\cite{galb} due to the large collecting area of the
technique, $\sim 5 \times 10^4$ m$^2$.  Only recently, however, has it
become possible to overcome the main factor limiting the sensitivities
of ground-based observatories which utilize this method. This is the
high rate of the cosmic ray (CR) background events. Introduction of
the two techniques, imaging and stereoscopy, pioneered by the Whipple
and HEGRA collaborations respectively ~\cite{Reynolds93,Aharonian93},
has made a drastic improvement in the suppression of the background,
and has made possible the reliable detection and spectroscopy of
several galactic and extragalactic sources
~\cite{Crab,Mrk421,Mrk501,Mrk501H}. Recent reviews of the development
of high energy ground-based $\gamma$-ray astronomy can be found
in~\cite{Ong98,MT}.

Methods for the efficient discrimination of photon and hadron
initiated events have been derived from the differences in the
intrinsic properties of the Cherenkov radiation from purely
electromagnetic and hadronic cascades. The most prominent of these is
the substantially wider distribution of the arrival directions of
photons in hadronic showers due to an additional angular scattering of
the secondary products during hadronic interactions. This effect has
been extensively studied in Monte-Carlo simulations of the cascade
development in the atmosphere~\cite{AP1,MH1}, and the rejection
criteria for the cosmic ray background has been derived in the form of
the so-called ``shape cut''~\cite{DFegan92,Aharonian97}.  Increased
angular resolution of ground-based $\gamma$-ray observatories together
with this means of identification of the primary photon are the most
important components of the recent discoveries and advances made in
the highest energy $\gamma$-ray astronomy.

The confidence of discovery of a $\gamma$-ray source in the case of a
background dominated data sample is defined as
\begin {eqnarray}
R = \frac {N _ {\mathrm{on}} -N _ {\mathrm{off}}}
{\sqrt {2N _ {\mathrm{off}}}},
\label{R_def}
\end {eqnarray}
where $N_{\mathrm{on}}$ is the number of detected events produced by a
hypothetic source for a given exposure, and $N_{\mathrm{off}}$ is the
number of
background events for the same observation time.  This quantity
reflects an excess of events due to the presence of a source measured
in units of background standard deviation shown in the denominator for
a Poisson distribution. It is generally accepted that the existence of
a source is established if $R \geq 5 $.

By application of the background rejection procedure one increases the confidence of the source
discovery due to elimination of the signal
 fluctuations connected with the background. The efficiency of the
 discrimination technique used to identify primary photons and
reject cosmic ray events is expressed then as
\begin {eqnarray}
\eta =\frac {\kappa_{\gamma}} {\sqrt {\kappa_{\mathrm{p}}}},
\label{eta}
\end {eqnarray}
where $ \kappa_{\gamma} $ and $ \kappa_{\mathrm{p}} $ are the probabilities of acceptance of
photon and background events after the application of selection criteria. Thus the $\eta$ factor
is an enhancement of $R$ indicating that $\eta^2$ times less exposure is required for a source
discovery when a new discrimination technique is applied. This is true in the case of background
dominated regime, when $N
_ {\mathrm{on}}-N _ {\mathrm{off}}\leq N _ {\mathrm{off}}$. On the contrary, in the
 statistics dominated regime ($N
_ {\mathrm{on}}-N _ {\mathrm{off}} >> N _ {\mathrm{off}}$), background rejection procedures with
$\kappa_{\mathrm{\gamma}}$ remarkably smaller than 1 should be avoided.

In this paper we explore a new discrimination technique that is
characterized by efficiency $\eta=1.5 - 2$ in addition to the already
existing methods of background suppression achieved by imposing shape
and orientation cuts. This technique is based on the differences in
intrinsic fluctuations of Cherenkov radiation produced in the pure
electromagnetic and hadronic cascades. We adapt the maximum likelihood
approach based on an analysis of the $\chi^2$ functional to establish
discrimination criteria to distinguish showers initiated by
$\gamma$-rays from those initiated by CR nuclei. This idea was induced
originally by the study of the mass composition of primary cosmic
radiation~\cite{AP2} where similar methods were proved to be very
successful.

\section{The intrinsic fluctuations of the image}

The imaging atmospheric Cherenkov telescope (IACT) consists of a set
of mirrors on a single mount with a two-dimensional array of
close-packed hexagonal photomultiplier tubes (pixels) in the focal
plane. Therefore information about an event is recorded in the form of
a two-dimensional image describing the distribution of the light
intensity in the focal plane. Ultimately, the Cherenkov light image is
a set of ADC counts corresponding to all pixels. But for simplicity we
deal with the number of photoelectrons emitted from the pixels'
photocathodes. Hence the Cherenkov
light intensity in the focal plane is characterized by the continuous
two dimensional distribution of photoelectron density, $\rho \lp x,y
\rp $. Each point $\lp x, y \rp$ on the focal plane corresponds to a
certain arrival direction of photons relative to the pointing of the
telescope mount.  The total number of photoelectrons (pes) in the
Cherenkov light image, $SIZE$, is given by
\begin{eqnarray*}
SIZE=\int{\rho\lp x,y \rp \, \d x \d y}.
\end{eqnarray*}
In our further calculations we make frequent use of the normalized to
unity distribution of photoelectrons

\begin{eqnarray}
 f\lp x,y \rp=SIZE^{-1}\rho\lp x,y \rp,
\label{norm_distr}
\end{eqnarray}
which is equivalent to the probability density function (pdf).  By
translation and rotation of the reference frame it is always possible
to choose a new coordinate system $(x,y)$ in which the first moments
of $f\lp x,y \rp$ are equal to zero and the matrix of second moments
is diagonalized. The diagonal elements
\begin{eqnarray*}
\int{x^2 f\lp x,y \rp \, \d x \d y} & = & LENGTH^2 \\
\int{y^2 f\lp x,y \rp \, \d x \d y} & = & WIDTH^2,
\end{eqnarray*}
which characterize the extent of the Cherenkov light distribution
along the major and minor axes of the image, are often used as the
image shape parameters.

Let us denote also the marginal distributions
\begin{eqnarray*}
 f_L\lp x \rp  & = & \int{f\lp x,y \rp \, \d y},\\
 f_W\lp y \rp  & = & \int{f\lp x,y \rp \, \d x},
\end{eqnarray*}
where indices $L$ and $W$ correspond to the major and minor axes of
the image respectively. We estimate $f_L\lp x \rp$ and $f_W\lp y \rp$
by the grid functions
\begin{eqnarray*}
 N_{L,k}=\int_{x_k}^{x_{k+1}}{f_L\lp x \rp \, \d x},
\qquad x_{k+1}=x_k+\Delta x; \\
 N_{W,k}=\int_{y_k}^{y_{k+1}}{f_W\lp y \rp \, \d y},
\qquad y_{k+1}=y_k+\Delta y;
\end{eqnarray*}
defined at $\{x_k, y_k; k=1,\ldots ,K\}$.

Suppose that we have two sets of shower images, initiated by
photons and cosmic rays, for which the mean values,
$\bar N_{t,k}^{\lp \gamma \rp}, \bar N_{t,k}^{\lp \mathrm{p} \rp}$,
and variances,
$\sigma^2_{N_{t,k}^{\lp \gamma \rp}},
\sigma^2_{N_{t,k}^{\lp \mathrm{p} \rp}}$ of the grid functions
are found for each $k$,\mbox{$t\in \{ L,W \} $}. For each image we
construct a pair of functionals $\chi^2_t$
\begin{eqnarray}
\chi^2_t=\frac{1}{K}\sum_{k=1}^{K}
{
 \frac
 {{\lp N_{t,k}- \bar N_{t,k}^{\lp \gamma \rp} \rp}^2}
 {\sigsq{\gamma}}
}
\end{eqnarray}
so that the mean values of $\chi^2_t$ for $\gamma$-showers and
background events are equal to:
\begin{eqnarray}
\overline{\chi_t^{2\lp \gamma \rp}}=1; \quad
\overline{\chi_t^{2\lp \mathrm{p} \rp}}=\frac{1}{K}\sum_{k=1}^{K}
{
 {\lb\frac{\sig{\mathrm{p}}}{\sig{\gamma}}\rb}^2
} + \frac{1}{K}\sum_{k=1}^{K}
{
 {\lb\frac{\bar N_{t,k}^{\lp \gamma \rp}-
\bar N_{t,k}^{\lp \mathrm{p} \rp}}{\sig{\gamma}}\rb}^2.
}
\label{2contrib}
\end{eqnarray}
The $\overline{\chi_t^{2\lp \mathrm{p} \rp}}$ consists of two
terms. The first term reflects the differences in the fluctuations
of the light intensity of images initiated by $\gamma$-rays and
those initiated by CR particles.
The second one characterizes the differences between the average
shapes of distributions of the photoelectron density in the images
of the two data sets.

Strickly speaking, two different effects influence the value of
the  ''fluctuation'' term of $\overline{\chi_t^{2\lp \mathrm{p} \rp}}$.
The first of them is connected with random
variations of the angular size of the image described, for example,
by WIDTH and LENGTH parameters. The second effect is
the irregularity of the light distribution inside the image itself.
We say in this case about the intrinsic fluctuations. Even in the
case of $\gamma$-ray and proton induced images having the same
angular size the first of them has a more smooth and regular
structure and, therefore, a smaller value of intrinsic fluctuations.
Large fluctuations in the number of secondary particles created during
the hadron multi-particle production is the main reason of
difference in intrinsic fluctuations of $\gamma$- and $p$-induced
images.

In this work we attempt to make use
of the contribution to $\overline{\chi_t^{2\lp \mathrm{p} \rp}}$
from the first (``fluctuation'')  term because  we expect that
the second term  (as well as the fluctuations connected
with variations of the angular size of the image) is rather small for images
which passed the image shape selection criteria.

Keeping in mind the argumentation presented above,
one can hope to use selection criteria of the form

\begin{eqnarray}
\chi^2_L<\tilde\chi^2_L, \qquad \chi^2_W<\tilde \chi^2_W
\label{single_xi_cond}
\end{eqnarray}
to reject CR-induced events and retain genuine photon-initiated
showers.  For a single telescope, two constants $\tilde\chi^2_L$ and
$\tilde\chi^2_W$ should be optimized to maximize the signal to noise
ratio (see formula (\ref{R_def})). For a system of telescopes a set
of similar criteria can be utilized
\begin{eqnarray}
 \chi^2_{L,i}<\tilde\chi^2_L, \qquad \chi^2_{W,i}<\tilde \chi^2_W,
 \qquad i=1,\ldots ,N_{\mathrm{trig}},
 \label{syst_xi_cond1}
\end{eqnarray}
in which condition (\ref{single_xi_cond}) is satisfied simultaneously
for all triggering telescopes. We consider a slightly more relaxed
background rejection criterion
\begin{eqnarray}
\frac{1}{N_{\mathrm{trig}}}\sum_{i=1}^{N_{\mathrm{trig}}}
{\chi^2_{L,i}}<\tilde\chi^2_L, \qquad
\frac{1}{N_{\mathrm{trig}}}\sum_{i=1}^{N_{\mathrm{trig}}}
{\chi^2_{W,i}}<\tilde\chi^2_W
\label{syst_xi_cond2}
\end{eqnarray}
which seems to retain more photon-initiated events and produce
a higher efficiency factor (\ref{eta}). In the case of a single
telescope, both criteria (\ref{single_xi_cond}) and
(\ref{syst_xi_cond2}) are identical.

The success of a selection criterion depends dramatically on how the
data sets of photons and CR nuclei have been selected. In order to
increase the selection efficiency one should consider the smaller
regions of the primary photon parameter phase space and formulate
selection criteria for each of them. We expect that the shower
parameters which affect background rejection most strongly are the
primary energy, $E$ and the impact parameter, $r$ (the shortest
distance between the telescope and the cascade core).  In order to
make the $\chi^{2\lp \gamma \rp}$ distribution narrower and thus
improve the discrimination ability of the method, we will evaluate the
dependence of the grid functions of photoelectron density on these
parameters
\begin{eqnarray}
 N_{t,k}=N_{t,k}\lp E, r \rp, \quad \bar N_{t,k}^{\lp \gamma \rp}
 =\bar N_{t,k}^{\lp \gamma \rp}\lp E, r \rp ,
 \quad \sig{\gamma}=\sig{\gamma}\lp E, r \rp.
 \label{E_R_dep}
\end{eqnarray}

We would
like to note that the $\chi^2$-based method has been used in
the past to deduce parameters of the primary
particle~\cite{Ulrich98,LeBohec98}. In our work, however, we do
not explicitly assume that the fluctuations in the images of
$\gamma$-rays are of a Poissonian nature.

\section {Data sets of shower images}

\begin {figure} [th]
\centering
\includegraphics [angle=-90, width=50mm]
 {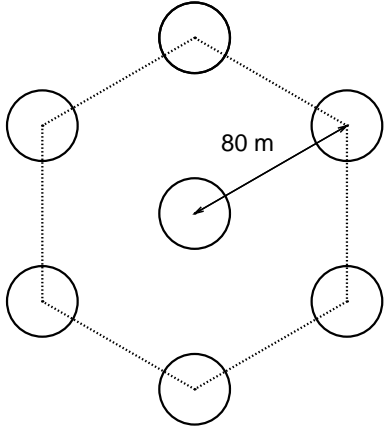}
\caption {Layout of VERITAS telescopes.}
\label {Veritas_tels}
\end {figure}
To investigate the discrimination efficiency of the method we have
simulated two data bases of photon and proton initiated showers. This
task has been accomplished by utilizing ALTAI computation
code~\cite{ALTAI} with input data summarized in
Table \ref{Veritas_config}. The geometrical layout of the VERITAS array
of telescopes is shown in Fig. \ref{Veritas_tels}.
\begin{table}[hb]
\caption {Basic parameters of VERITAS telescopes}.
\begin{tabular}{l r}      \hline
Number of telescopes & 7 \\ Altitude above the sea level, [m] & 1500 \\ Area of the mirror,
[m${}^2$] & 78 \\ Number of pixels & 547 \\ Pixel size, [degree] & 0.15 \\ Field of view,
[degree] & $\simeq$4.0 \\ \hline
\end{tabular}
\label{Veritas_config}
\end{table}
Although, both proton and photon induced cascades were simulated as
vertical showers, images of cosmic rays were also sampled
isotropically on the focal planes of the telescopes near the vertical
direction.  44,000 and 92,000 showers were simulated
for photons and protons respectively. The position of the shower core
was distributed uniformly inside a circle with radius of $300$ m
centered on the central telescope of the VERITAS array. A power law
differential energy spectra of primary particles was assumed with
exponent $-2.6$ for photons and $-2.75$ for protons. Photons were
sampled from the energy interval \mbox {$ 0.05 - 10 $ TeV} and protons
were distributed between \mbox {$ 0.1 - 20 $TeV}.

The contribution of the night sky background was taken to be one
photoelectron per pixel. We applied a rather severe cleaning algorithm
which insured that only bright images with light intensity well above
this background were included in our analysis. Shower images were
processed in two stages. First, all pixels with less than five
photoelectrons were excluded from further consideration. Then, only
pixels which contained more than nine photoelectrons or which had a
neighbor pixel with a nonzero content were assumed to be image
pixels. In addition, only images with $SIZE$ greater than 100
photoelectrons were accepted for the analysis. The telescope was
considered to be triggered when at least two pixels from an internal
zone of the multichannel camera (331 inner pixels) detected more than
20 photoelectrons. Finally, to ensure the high quality of images, only
telescopes with maximal light intensity of the image in the internal
zone were accepted.

\section {Determination of the impact parameter}

The efficiency factor, $\eta$, of the discrimination technique
increases if the dependence of the image photoelectron density on the
energy and the impact parameter of a shower is accounted for
(formula(\ref{E_R_dep})).  Therefore, it is necessary to have some
method of determining these quantities.  In this work we have
investigated three possibilities for estimating impact parameter of
the primary particle:
\begin {itemize}
\item VERITAS operates as a system of telescopes. The impact parameter
is determined by a simple geometrical event reconstruction method
(see, e.g.,~\cite{Aharonian97}).
\item VERITAS operates in a single telescope mode.
      \begin {itemize}
      \item The impact parameter  is determined on the basis of the
       $DIST$ (distance) parameter of the image.
      \item The impact parameter  is determined by means of the
        $ELLIPT$ (ellipticity) parameter of the image.
      \end {itemize}
\end{itemize}

Stereoscopic observation of an event by at least two telescopes allows
unambiguous geometrical reconstruction of the shower core location in
space.  If the number of telescopes which detected an event is larger
than two it is possible to increase the accuracy in determining
this parameter. In this work we use as an estimate of the shower core
location, a weighted average of the core coordinates derived from the
data of each pair of triggered telescopes~\cite{Ulrich98}.

\begin{figure}[ht]
\includegraphics[angle=-90, width=80mm,
keepaspectratio=true]{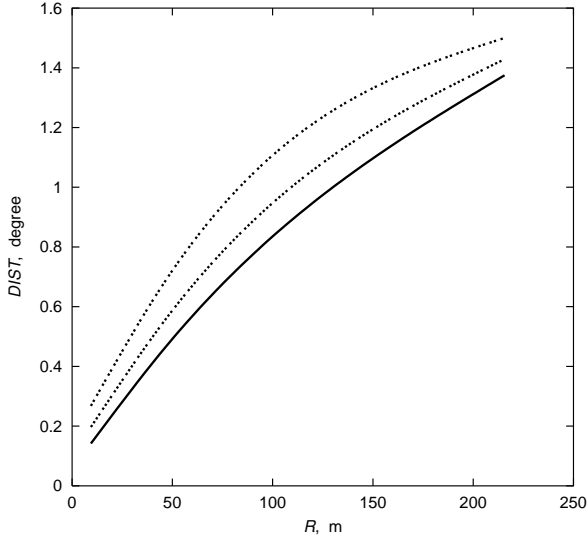}
\caption {Dependence of the $ DIST $ parameter on the impact parameter
for $\gamma$-showers. The curves are calculated for $SIZE$ intervals
with the centres at 220 (bottom curve), 840, 3200 pes.
(top curve) respectively.}
\label{dist_via_r}
\end{figure}

For a single telescope, determination of the impact parameter utilizes
its correlation with such characteristics of the image, as $ELLIPT$ or
$DIST$~\cite{AP3} defined as
\begin{eqnarray}
 ELLIPT=\frac{LENGTH}{WIDTH}-1, \qquad
 DIST=\sqrt{x_{\mathrm{c}}^2+y_{\mathrm{c}}^2},
\end{eqnarray}
where $x_{\mathrm{c}}$ and $y_{\mathrm{c}}$ are coordinates of
the image centre of gravity
in the reference frame with an origin at the source location.  Unlike
the $DIST$ parameter, $ELLIPT$ provides an opportunity for determining
the impact parameter not only for a point-like, but also for an
extended $\gamma$-ray sources when the arrival direction of a photon
is not known. To derive our estimate of impact parameter in the case
of a single telescope observations, we tabulated mean values of
$ELLIPT$ and $DIST$ for a set of values of impact parameter and image
$SIZE$. The required estimate of $r$ was then calculated by backward
interpolation. In \mbox {fig. \ref{dist_via_r}} we show the dependence
of mean $DIST$ on the impact parameter for several intervals of
$SIZE$. Larger values of $SIZE$ correspond to curves with larger
values of $DIST$.

\begin{figure}[ht]
\includegraphics[angle=-90, width=80mm,
keepaspectratio=true]{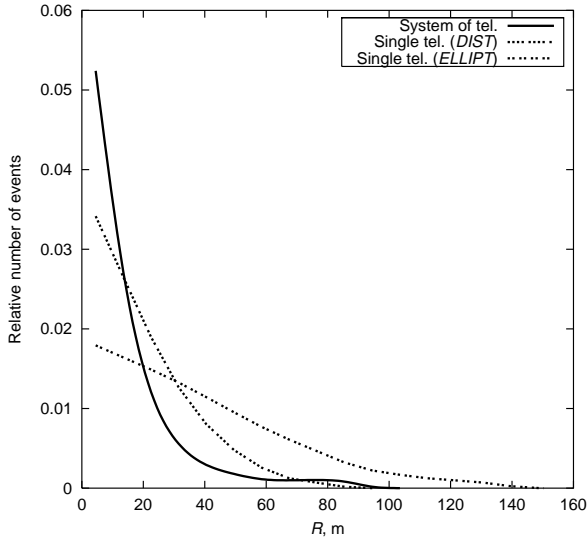}
\caption {The probability distribution for the error of determining
the $\gamma$-shower impact parameter. The $SIZE$ of the images is larger
than 150 pes.}
\label{r_dens}
\end{figure}

In \mbox {fig.\ref{r_dens}} and tab.\ref{r_err_integ} we show the
accuracy of determination of the impact parameter by three different
methods.

\begin{table}[hb]
\caption {The value of the impact parameter
determination error [m].  $50\%$ of showers have an error smaller than
the value presented in the table. The $SIZE$ of the images is larger
than 150 pes.}
\begin{tabular}{l c c}\hline
        &$\gamma\,       $& $p$           \\ \hline
System of telescopes           & 5               &32     \\
Single telescope ($DIST$)  & 11              &53         \\
Single telescope ($ELLIPT$)& 27              &68         \\ \hline
\end{tabular}
\label{r_err_integ}
\end{table}

The accuracy of determining the impact parameter depends strongly
on the lower bound of the $SIZE$ of images included in the
analysis.  The smaller the $SIZE$ of the shower image, the
stronger the effects of fluctuations in image photoelectron
density, and the stronger the influence of the night sky
background. A large difference between the errors of impact
parameter determination for photons and protons in the case of a
system of telescopes is due to the inherent differences in the
electron density distribution between the images of showers
initiated by these primaries.  While the distribution of Cherenkov
light is smooth and compact for $\gamma$-rays, for protons it is
fuzzy and more extended.
As a result the proton induced images
(and their basic parameters determining the reconstructed value of the
shower impact parameter) are more significantly distorted in the process of
the cleaning procedure  aimed at the elimination of the night
sky background and consisting in an exclusion from  consideration of
pixels with low magnitudes.

In the case of the photon primary the method
of impact parameter determination based on the $DIST$ parameter is
substantially more accurate than the one where the ellipticity is
in use. Perhaps, this can be explained by the larger fluctuations
of the $ELLIPT$ parameter, or in other words by weaker correlation
of $ELLIPT$ with shower core location.  The "ELLIPT" based method
indicates a substantially smaller error on the impact parameter
determination for $\gamma$-induced air showers relative to proton
initiated ones, which can also be explained by the larger
fluctuations of the light intensity in CR-induced images. For
isotropically distributed cosmic rays one should not expect any
correlation between the impact parameter of the showers and the
$DIST$ value. The error on the determination of the shower core
location of $53$~m, found in this case, is finite because of the
limited telescope field of view and the high directionality of the
distribution of Cherenkov photons from atmospheric cascades.

\section{ Event selection based on the shape parameters}

\begin{figure}[ht]
\includegraphics[angle = -90, width=80mm, keepaspectratio=true]{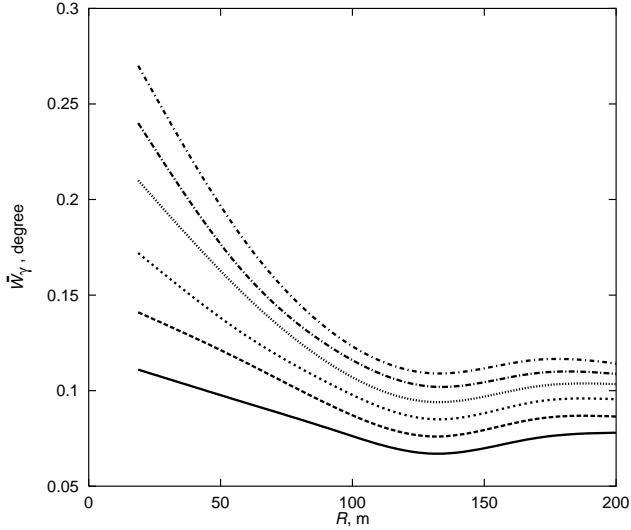}
\caption{$\bar W^\gamma=f(SIZE, r)$. The curves are calculated for
$SIZE$ intervals with the centres at 220 (bottom curve), 430, 840, 1600,
3200, 6200 pes (top curve), respectively}
\label{wsfig}
\end{figure}

To estimate the discrimination efficiency of this new technique, we
applied a preliminary event selection on the basis of the shape
parameters ($WIDTH$ and $LENGTH$) thereby excluding any possible
correlation between the traditional and proposed event classification
methods.  In this work we adapted a selection criterion in the form
\begin{eqnarray}
\frac{WIDTH}{\bar W^{\lp \gamma \rp }\lp SIZE,r \rp}
\leq \tilde W_{\mathrm{sc}}, \quad i=1, \ldots , N_{\mathrm{trig}}
 \label{wscsel}
\end{eqnarray}
where $\bar W^{\lp \gamma \rp} \lp SIZE, r \rp$ is the mean value
of $WIDTH$ for $\gamma$-showers detected with a given image $SIZE$
and estimated impact parameter, $r$, (fig. \ref{wsfig}).
Application of this criterion, commonly known as the scaled width
method~\cite{Aharonian97}, turns out to be particularly effective
as it accounts for the changes of image $WIDTH$ caused by the
differences in primary photon energy and the possible variations
in the shower core location.
We excluded $LENGTH$ as a discrimination parameter while using
$W_{\mathrm{sc}}$; two-dimensional discrimination on the basis
of $LENGTH$ and $WIDTH$ yields only negligible increase in
$\eta$~\cite{Aharonian97}.

\begin{figure}[ht]
\includegraphics[angle=-90,width=80mm,
keepaspectratio=true] {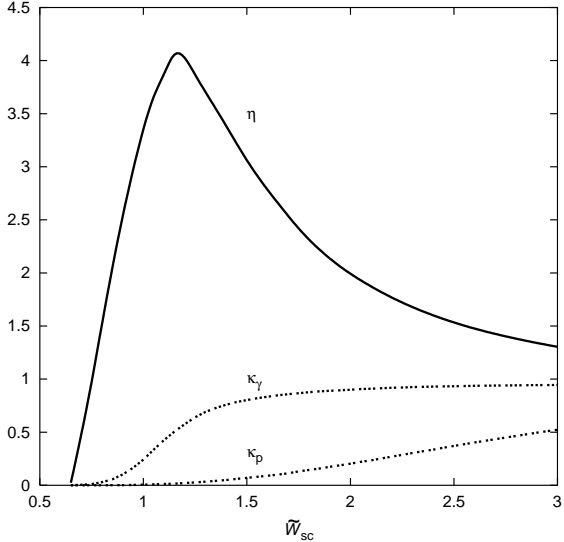}
\caption{ Dependence of background discrimination efficiency $\eta$,
true classification probability ( $\kappa_\gamma$) of $\gamma$-rays,
and false classification probability ($\kappa_{\mathrm{p}}$) of protons
as functions of the $W_{\mathrm{sc}}$ cut. In this example, VERITAS
operates as an array of telescopes.}
\label{wselfig}
\end{figure}

\begin{figure}[ht]
\includegraphics[angle=-90, width=80mm,
keepaspectratio=true]{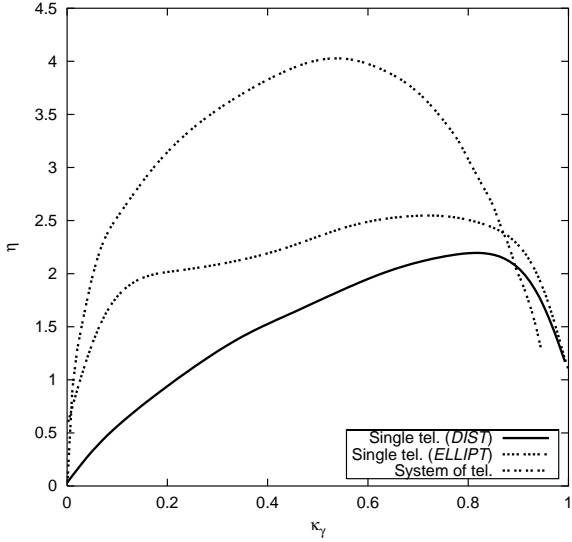}
\caption{ Dependence of the background discrimination efficiency $\eta$
on the true classification probability $\kappa_\gamma$ of photons.}
\label{wseleffig}
\end{figure}

In fig.~\ref{wselfig} the dependence of background discrimination
efficiency on $\tilde W_{\mathrm{sc}}$ cut (\ref{wscsel}) is shown. The
dependence of $\eta$ on the true classification probability of
$\gamma$-rays is shown, for different event reconstruction methods, in
fig.~\ref{wseleffig}.

\section{Optimization of the discrimination technique}

\begin{figure}[ht]
\includegraphics[angle=-90,width=80mm,
keepaspectratio=true]{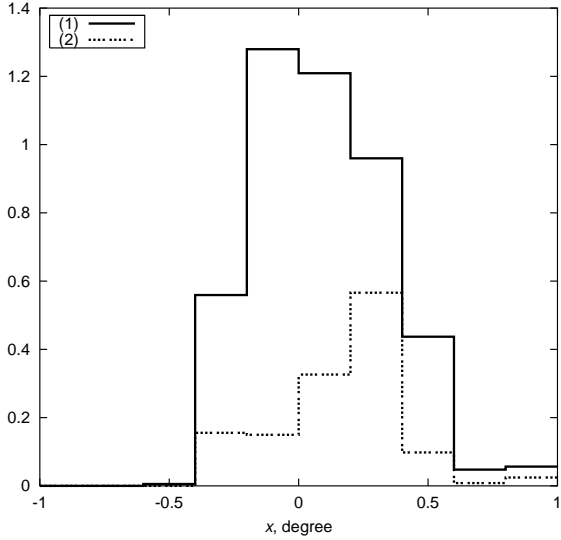}
\caption{ $\bar N_{L,k}^{\lp \gamma \rp}/
 \bar N_{L,k}^{\lp \mathrm{p}\rp}$ (1)
and $\sigma_{N_{L,k}^{\lp \gamma \rp}} /
\sigma_{N_{L,k}^{\lp \mathrm{p}\rp}}$(2)
for events surviving the scaled width cut ($W_{\mathrm{sc}}=1.4$,
$DIST$-approach). A single VERITAS telescope is considered.
Bin size $\Delta x=0.2^\circ$.}
\label{xiprof}
\end{figure}

Fig.~\ref{xiprof} depicts two histograms: the ratio of the mean
photoelectron density along the major axis $(t=L)$ for images
initiated by photons and protons (1), and the ratio of their
standard deviations (2). The $SIZE$ parameter was limited within
interval from $100$ to $200$ photoelectrons, and the actual shower
impact parameter was restricted to the $50 - 75$ m interval.
The values of
these intervals were choosen as small as the statistics allows
in order to reduce the  fluctuations connected with random variations of
the primary energy  and of the  impact  parameter.

It can be seen that for the given conditions, the contribution
to the $\chi^2$ functional (Eq.~\ref{2contrib}) from the first,
``fluctuation'', term exceeds the second, ``mean difference'',
term by almost a factor of ten.
This confirms our assumption about the potential of the
Cherenkov light intensity fluctuations for discriminating
proton and $\gamma$-induced images.

The bin sizes, $ \Delta x = 0.2 ^\circ $ and $ \Delta y= 0.06
^\circ $, used to construct the grid photoelectron density
functions, were optimized to achieve the highest discrimination
efficiencies. They are chosen as a balance between the necessity
to resolve features in the shower images on the scales smaller
than $LENGTH$ and $WIDTH$ and, at the same time, not to make
fluctuations in the bins dominated by Poisson statistics.
In addition, a special analysis has shown that the
$\chi^2$-technique exhibits no essential sensitivity
to the bin size if the value of this size does not
exceed considerably the  pixel size ($0.15^o$) of the VERITAS telescopes.

\begin{figure}[ht]
\includegraphics[angle=-90, width=70mm, keepaspectratio=true]
{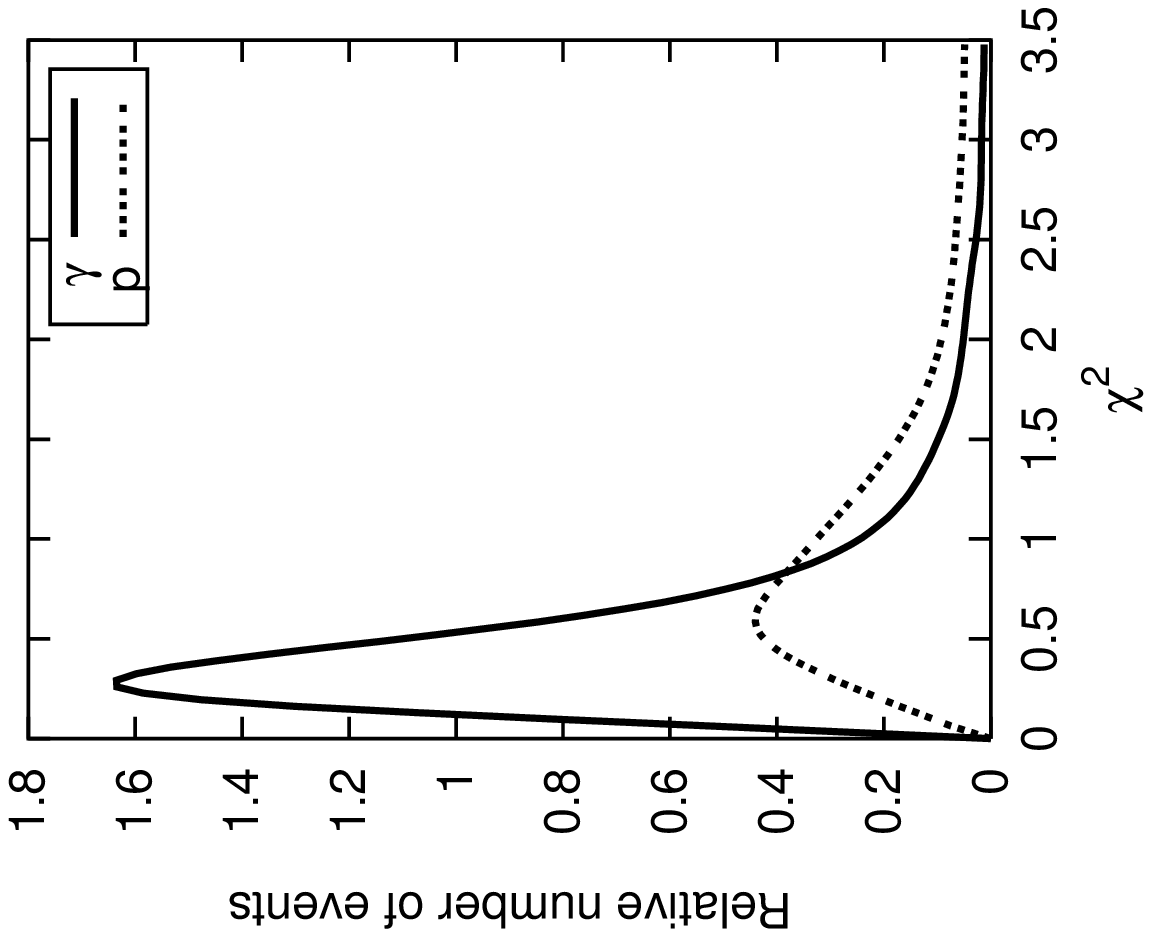} \hfill
\includegraphics[angle=-90, width=70mm,
keepaspectratio=true]{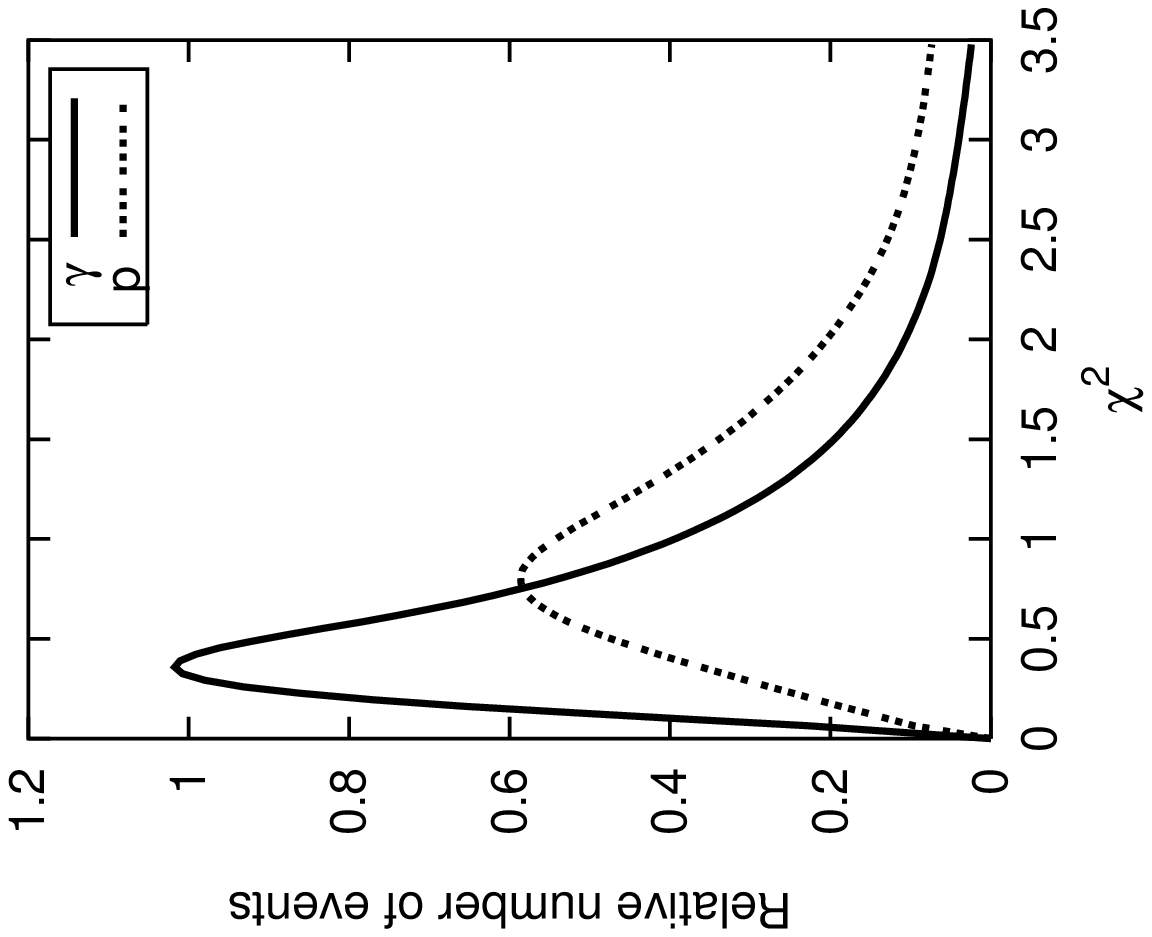}
\caption{Probability distributions for $\chi^2_L$ (on the left) and
$\chi^2_W$ (on the right) for a single VERITAS telescope
after $W_{sc}=1.3$ cut is applied. Determination of the impact parameter
is based on $DIST$ parameter. }
\label{xidistr}
\end{figure}

The comparison of the probability distributions $\chi^2_L$ and
$\chi^2_W$ for $\gamma$ and $p$-induced events, for a single
telescope, is shown in fig.~\ref{xidistr}.
Tab.~\ref{sysxisel} summarizes the photon selection efficiency after
application of the scaled width cut, ($\kappa_{\gamma W}, \eta_W$),
and after discrimination of background using the $\chi^2_L $ and $
\chi^2_W $ parameters (eqn.~\ref{syst_xi_cond2}), (
$\kappa_{\gamma\chi}, \eta_{\chi}$).  These calculations were
performed for the full VERITAS array. The first column indicates the
value of $W_{\mathrm{sc}}$ cut. The total discrimination effect is
denoted as ($\kappa_{\gamma \mathrm{tot}}, \eta_{\mathrm{tot}}$).
$\kappa_{\gamma\chi}$ corresponds
to the value of $\chi^2_L $ and $ \chi^2_W $ cuts at which
discrimination efficiency factor, $\eta$, is maximal provided that
$\kappa_{\gamma\chi} \geq 0.7$.
\begin{table}[hbt]
\caption{ Summary of the background rejection efficiency for VERITAS. }
\begin{tabular}{c c c c c c c}\hline
$\tilde W_{\mathrm{sc}}$&$\kappa_{\gamma W}$&$\eta_W$&
$\kappa_{\gamma\chi}$&$\eta_{\chi}$&
$\kappa_{\gamma \mathrm{tot}}$&$\eta_{\mathrm{tot}}$\\ \hline
          1.15 &  0.52&  4.11&   0.81&    1.41&  0.42&   5.80\\
          1.20 &  0.60&  4.08&   0.74&    1.43&  0.44&   5.83\\
          1.25 &  0.67&  3.90&   0.79&    1.58&  0.53&   6.16\\
          1.3 &  0.72&  3.87&   0.82&    1.66&  0.59&   6.42\\
          1.4 &  0.79&  3.55&   0.70&    1.84&  0.55&   6.53\\ \hline
\end{tabular}
\label{sysxisel}
\end{table}

It is evident from the table that the largest value of the total
efficiency factor, $\eta_{\mathrm{tot}}$, is achieved for
$\tilde W_{\mathrm{sc}}$
larger than the value which provides maximum discrimination
power of photon selection based solely on the scaled width cut.
This is also correct for event reconstruction
using a single telescope. Selection criterion utilizing only the
$\chi^2_L$ cut demonstrates a discrimination efficiency factor
$ \simeq 80-90 \% $ of the one obtained with the use of
both  $ \chi^2 $ cuts. An increase of $\tilde W_{\mathrm{sc}}$ provides
a moderate growth of photon discrimination efficiency when
$\chi^2_W$ is sequentially applied, while the effect of a $\chi^2_L$
selection depends very weakly on the scaled width cut.

\begin{figure}[ht]
\includegraphics[angle=-90,width=80mm,
keepaspectratio=true]{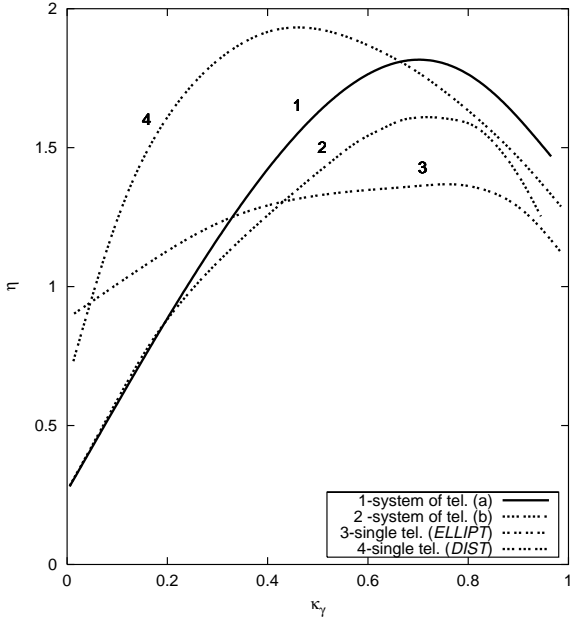}
\caption{Dependence of $\chi^2$ selection efficiency
$\eta\lp\kappa_\gamma\rp$ for different event reconstruction methods.
For array of telescopes, curve (a) corresponds to $\chi^2$ selection
criterion (\ref{syst_xi_cond2}), and curve (b) is obtained for criterion
(\ref{single_xi_cond}) applied to each telescope participating in
the VERITAS trigger.}
\label{xiefdep}
\end{figure}

The plots of discrimination efficiency of $\chi^2$ technique as a
function of true photon classification probability,
$\kappa_{\gamma}$, are shown in \mbox{fig.~\ref{xiefdep}} for various
VERITAS event reconstruction methods. The scaled width cut was
optimized to provide maximal total discrimination efficiency when
$\kappa_{\gamma}$ is fixed.  Curves (1) and (2) in this figure compare
the factor $\eta$ for a $\chi^2$ cut in the form given in
(\ref{syst_xi_cond2}) and (\ref{single_xi_cond}) when the latter is
applied to each VERITAS telescope in the event trigger.
It can be seen that the discrimination efficiency of photon selection
is higher for the case which uses the mean value of $\chi^2$.

In tab.~\ref{sdep} we present the true photon classification
probability and discrimination efficiency of the $\chi^2$ technique
when various lower bounds on image $SIZE$ are included in the
analysis.  The dependence of these quantities on the lower limit of
the estimated energy of the primary particle is shown in the
tab.~\ref{edep}.  For both tables we tuned the scaled width cut to
optimize the total efficiency factor. Primary energy was estimated by
backward interpolation with respect to image $SIZE$ and estimated
impact parameter, $r$, using simulated data base of photon induced
showers.
\begin{table}[hbt]
\caption{ Discrimination efficiency of
$\chi^2$ photon selection technique ( $\tilde W_{\mathrm{sc}}$ cut is
tuned to maximize $\eta_{\mathrm{tot}}$). Different image $SIZE$
lower bound for analyzed events is shown.}
\begin{tabular}{l c c c c c c c c c c c c}\hline
&\multicolumn{4}{c}{System} &\multicolumn{4}{c}{Single($DIST$)}&
\multicolumn{4}{c}{Single($ELLIPT$)}\\ \hline

$S$  &100 &150 &200 &300 &100 &150 &200 &300 &100 &150 &200
  &300       \\
$\kappa_\gamma$ &0.75 &0.70 &0.71 &0.71 &0.73 &0.70 &0.73 &0.70 &0.79
&0.71   &0.70   &0.70\\
$\eta_{\chi^2}$ &1.50 &1.84 &1.94 &2.14 &1.43 &1.81 &2.34 &4.11
&1.22 &1.37 &1.56  &1.95 \\ \hline
\end{tabular}
\label{sdep}
\end{table}
\begin{table}[hbt]
\caption{ Discrimination efficiency of
$\chi^2$ photon selection technique as a function of lower bound
on event estimated energy to be included into the analysis.}
\begin{tabular}{l  c c c c  c c c c  c c c c}\hline

 &\multicolumn{4}{c}{System} &\multicolumn{4}{c}{Single ($DIST$)}&
 \multicolumn{4}{c}{Single ($ELLIPT$)}\\ \hline

$E_{\mathrm{est}}$, [TeV] &0.2 &0.3 &0.5 &1 &0.2 &0.3 &0.5 &1 &0.2 &0.3 &0.5
&1  \\
$\kappa_\gamma$ &0.72 &0.73 &0.70 &0.71 &0.70 &0.70 &0.71 &0.70 &0.72
&0.71   &0.71   &0.70   \\
$\eta_{\chi^2}$ &1.91 &2.00 &2.19 &3.33 &1.62 &1.97 &2.31 &3.92 &1.52
&1.65   &1.76   &2.47   \\ \hline
\end{tabular}
\label{edep}
\end{table}

\section {Conclusion}
In this work we have examined the feasibility of improving photon
selection using differences in the fluctuations of the Cherenkov light
distribution between the images of $\gamma$ and $p$ -initiated
showers. Three different event reconstruction methods which are likely
be utilized in the next generation of ground-based $\gamma$-ray
observatories, such as VERITAS, have been studied. Two of them are
applicable to observations with a single telescope and one uses the
stereoscopic capability of VERITAS array.

Optimization of our technique indicates that the proposed method
provides an additional increase of the total discrimination
efficiency by a factor $\ge 1.5 - 2.0$ retaining the survival
probability of $\gamma$-ray initiated events at the level larger
than $0.7$. That is true for both the telescope array and the
single telescope (with estimation of impact parameter by the
$DIST$-based method).  The
achievable value of the total discrimination efficiency is equal
to 6.5 and 3.6 respectively. When a single telescope event is
reconstructed utilizing the $ELLIPT$ parameter, our approach
provides an additional discrimination efficiency factor $\ge 1.4$.
This is particularly important for a detection of extended sources
or sources whose position has not been accurately identified . The
discrimination efficiency of the proposed method increases rapidly
with the energy of the primary photon.

\begin{ack}
A.V. Plyasheshnikov thanks the Harvard-Smithsonian Center for
Astrophysics for the opportunity to visit the Whipple Observatory.
We acknowledge the help of Deirdre Horan and Stephen Fegan
in preparation of the manuscript. This work was partially supported
by the US Department of Energy.
\end{ack}


\end{document}